\documentclass[aps,preprintnumbers,superscriptaddress,nofootinbib,prl,twocolumn,floatfix]{revtex4-2}
\pdfoutput=1

\usepackage{multirow,array}
\usepackage{amsmath,amssymb}
\usepackage{graphicx}
\usepackage{color}
\usepackage{url}
\usepackage{slashed}
\usepackage{footmisc}
\usepackage{hyperref}
\usepackage{braket}
\usepackage{colortbl}
\usepackage{orcidlink}
\usepackage{physics}
\usepackage{mathrsfs}
\usepackage{comment}
\usepackage[normalem]{ulem}
\usepackage[T1]{fontenc}

\newcommand{\bea}{\begin{eqnarray}}
\newcommand{\eea}{\end{eqnarray}}
\newcommand{\be}{\begin{equation}}
\newcommand{\ee}{\end{equation}}
\newcommand{\bma}{\begin{pmatrix}}
\newcommand{\ema}{\end{pmatrix}}

\definecolor{nicered}{rgb}{0.5,0.,0.}
\definecolor{nicegreen}{rgb}{0.,0.5,0.}
\definecolor{niceblue}{rgb}{0.,0.,0.5}
\hypersetup{colorlinks,citecolor=nicegreen,linkcolor=nicered,urlcolor=niceblue}

\begin{document}

\title{Probing \boldmath{$\tau$} lepton dipole moments at future lepton colliders}

\author{Dario Buttazzo\orcidlink{0000-0002-4678-7975}}
\affiliation{Istituto Nazionale di Fisica Nucleare, Sezione di Pisa, I-56127 Pisa, Italy}
\author{Gabriele Levati\orcidlink{0009-0000-7693-2152}}
\affiliation{Albert Einstein Center for Fundamental Physics, Institute for
Theoretical Physics, University of Bern, Sidlerstrasse 5, 3012 Bern,
Switzerland}
\author{Yang Ma\orcidlink{0000-0002-9419-6598}}
\affiliation{Centre for Cosmology, Particle Physics and Phenomenology (CP3),
UCLouvain, 1348 Louvain-la-Neuve, Belgium}
\author{Fabio Maltoni\orcidlink{0000-0003-4890-0676}}
\affiliation{Centre for Cosmology, Particle Physics and Phenomenology (CP3),
UCLouvain, 1348 Louvain-la-Neuve, Belgium}
\affiliation{Dipartimento di Fisica e Astronomia, Università di Bologna and INFN,
Sezione di Bologna, via Irnerio 46, 40126 Bologna, Italy}
\affiliation{CERN, Theoretical Physics Department, CH-1211 Geneva 23, Switzerland}
\author{Paride Paradisi\orcidlink{0000-0003-0556-7797}}
\affiliation{Dipartimento di Fisica e Astronomia `G. Galilei', Universit\`a di Padova, Italy}
\affiliation{Istituto Nazionale di Fisica Nucleare, Sezione di Padova, I--35131 Padova, Italy}
\author{ZeQiang Wang\orcidlink{0000-0002-1643-2727}}
\affiliation{Centre for Cosmology, Particle Physics and Phenomenology (CP3),
UCLouvain, 1348 Louvain-la-Neuve, Belgium}

\preprint{IRMP-CP3-26-07, CERN-TH-2026-089, COMETA-2026-06}

\begin{abstract}
Electron and muon dipole moments provide stringent tests of the Standard Model, whereas those of the $\tau$ remain weakly constrained. We investigate the sensitivities to $\tau$ dipoles at FCC-ee and a multi-TeV muon collider through $\ell^+\ell^- \to \tau^+\tau^-$ ($\ell=e,\mu$), associated Higgs production $\mu^+\mu^- \to \tau^+\tau^- H$, radiative Higgs decays $H \to \tau^+\tau^-\gamma$, and vector-boson scattering $\ell^+\ell^- \to \ell^+\ell^-\tau^+\tau^-$ and $\mu^+\mu^- \to \bar\nu\nu\tau^+\tau^-$. 
At FCC-ee, the unprecedented Tera-$Z$ luminosity enables SM--NP interference to dominate the sensitivity to magnetic dipole interactions in direct $\tau$-pair production, unlike at LEP-II, where quadratic NP effects dominated. 
Photon--photon collisions yield the strongest individual bound on the electromagnetic anomalous magnetic moment. Direct production also probes the weak dipoles, potentially offering stronger constraints in correlated NP scenarios. At a multi-TeV muon collider, associated Higgs production $\mu^+\mu^-\to\tau^+\tau^-H$ provides a complementary probe through the energy growth of the quadratic NP contribution. Together, these channels can improve existing bounds by several orders of magnitude.
\end{abstract}

\maketitle

\section{I. Introduction} 

Since the discovery of the Higgs boson at the LHC, no direct evidence for new physics (NP) has emerged at the TeV scale. Inspired by historical precedents in high energy physics where new particles first manifested through virtual effects, the precision frontier has become a primary avenue to probe NP dynamics via deviations from high-precision Standard Model (SM) predictions.

Lepton dipole moments provide sensitive indirect probes of NP. The electron and muon anomalous magnetic moments $a_{\ell}$ ($\ell=e,\mu$) are measured with high precision~\cite{Fan:2022eto,Muong-2:2023cdq,Muong-2:2025xyk}. Comparable theoretical accuracy is essential to maximize sensitivity to NP: for $a_e$ the dominant uncertainty stems from the fine-structure constant~\cite{Morel:2020dww,Parker:2018vye}, while for $a_\mu$ it arises from hadronic vacuum polarization~\cite{Aliberti:2025beg}. 
The current comparison between theory and experiment shows no significant discrepancy~\cite{Aliberti:2025beg},
motivating further reduction of theoretical errors~\cite{Hertzog:2025ssc,Aliberti:2025beg} to fully exploit the experimental sensitivity.
In contrast, the electron and muon electric dipole moments $d_{\ell}$ are negligible in the SM; any signal below current bounds~\cite{Roussy:2022cmp,Muong-2:2008ebm} would constitute clear evidence of CP-violating NP.

The $\tau$ lepton case is different. Its short lifetime prevents direct dipole measurements in external fields; $a_\tau$ and $d_\tau$ must instead be inferred from $\tau$-pair production. 
The SM prediction, $a_\tau^{\rm SM}=117\,717.1(4.0)\times10^{-8}$~\cite{Eidelman:2007sb,Eidelman:2016aih,DiLuzio:2024sps,Hoferichter:2025fea,Wittig:2025azm}, has an uncertainty well below foreseeable experimental sensitivity. These measurements uniquely probe NP with enhanced third-generation couplings, as motivated by models addressing the SM flavor puzzle~\cite{Barbieri:1995uv,Barbieri:1996ae,Barbieri:1996ww,Barbieri:1997tu,Kagan:2009bn,Barbieri:2012uh,Panico:2015jxa,Panico:2016ull,Barbieri:2012tu,Glioti:2024hye,Bordone:2017bld,Fuentes-Martin:2022xnb,Davighi:2023iks}.

NP effects on $\tau$ dipoles at a scale $\Lambda$ above the electroweak scale admit an effective field theory (EFT) description through gauge-invariant higher-dimensional operators~\cite{Buchmuller:1985jz}. 
In this framework, the DELPHI Collaboration at LEP-II obtained the 95\% C.L. limits $-0.052<a_\tau<0.013$ and $|d_\tau|<3.7\times10^{-16}\,e\,\mathrm{cm}$ from photon-photon collisions $\gamma\gamma\to\tau^+\tau^-$~\cite{DELPHI:2003nah}. 
At the LHC, recent measurements by ATLAS~\cite{ATLAS:2025oiy} and CMS~\cite{CMS:2024qjo} have improved these bounds, resulting in $a_\tau\in[-2.4,4.7]\times10^{-3}$ and $|d_\tau|<2.0\times10^{-17}\,e\,\mathrm{cm}$ at 95\% C.L.~\cite{ATLAS:2025oiy}. 
The High-Luminosity LHC is projected to reach $a_\tau\sim\mathcal{O}(10^{-3})$ and $|d_\tau|\sim\mathcal{O}(10^{-17})\,e\,\mathrm{cm}$,
therefore providing only a modest improvement over current bounds.

The strongest constraint on the $\tau$ electric dipole moment, $d_\tau\in[-1.85,0.61]\times10^{-17}\,e\,\mathrm{cm}$ at 95\% C.L., comes from Belle~\cite{Belle:2021ybo}, while no comparable bound 
on $a_\tau$ is currently available from $B$ factories.
Asymmetry-based strategies in $e^+e^-\to\tau^+\tau^-$~\cite{Bernabeu:2006wf,Bernabeu:2007rr,Bernabeu:2008ii} at Belle II target sensitivities of $|a_\tau|\lesssim10^{-5}$ and $|d_\tau|\lesssim10^{-19}\,e\,\mathrm{cm}$ with an integrated luminosity of $50\,\mathrm{ab}^{-1}$~\cite{Crivellin:2021spu,USBelleIIGroup:2022qro,Aihara:2024zds,Gogniat:2025eom,Hoferichter:2025zjp,Hoferichter:2025ijh,Belle-II:2018jsg, Gogniat:2026zvf}.

In this Letter, we examine the prospects for measuring the $\tau$ dipole moments at future high-energy lepton colliders, focusing on the $e^+e^-$ Future Circular Collider (FCC-ee)~\cite{FCC:2018evy,Bernardi:2022hny,FCC:2025lpp,Altmann:2025feg} and a multi-TeV Muon Collider 
($\mu C$)~\cite{MuonCollider:2022xlm,Aime:2022flm,Black:2022cth,Accettura:2023ked,Delahaye:2019omf,Bartosik:2020xwr,Schulte:2021hgo,Long:2020wfp,MuonCollider:2022nsa,MuonCollider:2022ded,MuonCollider:2022glg,InternationalMuonCollider:2025sys}. 
These facilities offer complementary strategies to probe $\tau$ dipole moments. 
At FCC-ee, the unprecedented Tera-Z luminosity enables SM--NP interference to dominate the sensitivity to magnetic dipole interactions in direct $e^+e^-\to\tau^+\tau^-$ production, unlike in the quadratic-dominated LEP-II regime. The $Z$-pole scan further helps disentangle electromagnetic and weak dipole interactions, while photon--photon collisions, $\gamma\gamma\to\tau^+\tau^-$, exploit low-invariant-mass subprocesses where interference effects are comparatively enhanced~\cite{Buttazzo:2026ulg}. At a multi-TeV $\mu C$, instead, the energy growth of quadratic NP contributions provides a complementary mechanism to probe $\tau$ dipole moments~\cite{Buttazzo:2026jfu}.

In addition to $\mu^+\mu^-\to\tau^+\tau^-$, we investigate vector-boson-scattering (VBS) processes, $\mu^+\mu^-\to\mu^+\mu^-\tau^+\tau^-$ and $\mu^+\mu^-\to\bar\nu\nu\tau^+\tau^-$, which probe $\tau$ dipole interactions through electroweak gauge-boson exchange. We also consider $\mu^+\mu^-\to\tau^+\tau^-H$, where the dipole operators contribute through the $H\tau^+\tau^-\gamma^*$ vertex. Although this channel is phase-space suppressed, the stronger energy dependence of its NP contribution can enhance its sensitivity at multi-TeV energies.

Finally, both FCC-ee and a $\mu$C can also operate as Higgs factories.
Therefore, the rare decay $H\to\tau^+\tau^-\gamma$~\cite{Buttazzo:2020ibd} provides an additional probe of the dipole operators through the same $H\tau^+\tau^-\gamma$ interaction. Although this channel is expected to yield weaker sensitivity than the direct production channels, it offers a complementary probe of $\tau$ dipole moments through Higgs decays.

\section{II. tau dipole moments in the SMEFT}
%
For $E\ll\Lambda$, we consider the leptonic dipole operators invariant under $SU(3)_c\otimes SU(2)_L\otimes U(1)_Y$~\cite{Buchmuller:1985jz,Grzadkowski:2010es}:

\begin{align}
\!\!\mathscr{L} &\supset 
\bar\ell_L \sigma^{\mu\nu}e_{R}
\left(
\frac{{\mathcal C}_{\ell B}}{\Lambda^2} 
\varphi B_{\mu\nu} + 
\frac{{\mathcal C}_{\ell W}}{\Lambda^2}
\tau^I \varphi W_{\mu\nu}^I 
\right)
\!+ {\rm h.c.}
\label{eq:L_SMEFT}
\end{align} 
%
where $\varphi$ contains the Higgs field $H$ and its vacuum expectation value $v=246\,\mathrm{GeV}$, while $B_{\mu\nu}$ and $W^I_{\mu\nu}$ are the $U(1)_Y$ and $SU(2)_L$ field-strength tensors.

Perturbative unitarity constrains the energy range where the EFT remains perturbatively reliable. For dipole operators, the bounds are~\cite{Allwicher:2021jkr,Bresciani:2026acy}
\begin{align}
s \,\frac{|{\mathcal C}_{\ell B}|}{\Lambda^2} < 4\pi\,, 
\qquad
s \,\frac{|{\mathcal C}_{\ell W}|}{\Lambda^2}  < 4\pi\, \sqrt{2/3}\,.
\label{eq:perturbative}
\end{align}
After electroweak symmetry breaking, Eq.~\eqref{eq:L_SMEFT} generates tree-level contributions to electromagnetic dipole moments:
\begin{align}
\Delta a_\tau  \simeq \frac{4m_\tau v}{e\sqrt{2}\Lambda^2} \, 
{\rm Re}\, {\mathcal C}_{\tau\gamma}~,\qquad
d_\tau  \simeq \frac{2 v}{\sqrt{2}\Lambda^2} \, 
{\rm Im}\, {\mathcal C}_{\tau\gamma}\,,
\label{eq:Delta_a_ell}
\end{align}
%
where ${\mathcal C}_{\tau\gamma}=c_W {\mathcal C}_{\tau B}-s_W {\mathcal C}_{\tau W}$, with $s_W$ ($c_W$) the sine (cosine) of the weak mixing angle. Equation~\eqref{eq:L_SMEFT} also induces neutral weak dipole moments, $\Delta a_\tau^Z$ and $d_\tau^Z$, obtained from Eq.~\eqref{eq:Delta_a_ell} by replacing ${\mathcal C}_{\tau\gamma}$ with ${\mathcal C}_{\tau Z}=-s_W{\mathcal C}_{\tau B}-c_W{\mathcal C}_{\tau W}$.
To illustrate the NP scales probed by $\Delta a_\tau$, Eq.~\eqref{eq:Delta_a_ell} gives
\begin{align}
\Delta a_\tau & \approx 
3.7\times 10^{-5} \left(\frac{10 \,{\rm TeV}}{\Lambda}\right)^{2}
{\rm Re}\, {\mathcal C}_{\tau\gamma} \,,
\end{align}
where ${\mathcal C}_{\tau\gamma}$ is evaluated at $m_\tau$, including one-loop renormalization effects~\cite{Jenkins:2013zja,Jenkins:2013wua,Alonso:2013hga,Jenkins:2017dyc,Aebischer:2021uvt}.

\medskip

These estimates imply:
\begin{itemize}
\item $\Delta a_\tau\sim\mathcal O(10^{-4})$ can be reached for $\Lambda\approx10\,\mathrm{TeV}$ and ${\rm Re}\,{\mathcal C}_{\tau\gamma}\sim1$, as expected for a strongly coupled NP sector with ${\rm Re}\,{\mathcal C}_{\tau\gamma}\sim g_{\rm NP}^2/(16\pi^2)\sim1$. This requires a chiral enhancement $v/m_\tau$ relative to the naive scaling $\Delta a_\tau\propto m_\tau^2$~\cite{Giudice:2012ms}.
\item For weakly coupled NP ($g_{\rm NP}\sim1$) with chiral enhancement, $\Delta a_\tau\sim10^{-4}$ can arise at $\Lambda\approx1\,\mathrm{TeV}$.
\item Without chiral enhancement, weakly coupled NP can generate $\Delta a_\tau\sim10^{-4}$ only for $\Lambda\lesssim v$, strongly disfavored by direct LEP and LHC searches.~\footnote{Effects from a very light NP sector ($\Lambda\lesssim1$ GeV) feebly coupled to SM particles remain allowed~\cite{Marciano:2016yhf,Alda:2024cxn,Hoferichter:2025zjp,Hoferichter:2025ijh}.}
\end{itemize}

\begin{figure}[ht]
\includegraphics[width=0.15\textwidth]{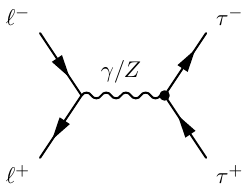}\qquad\qquad
\includegraphics[width=0.15\textwidth]{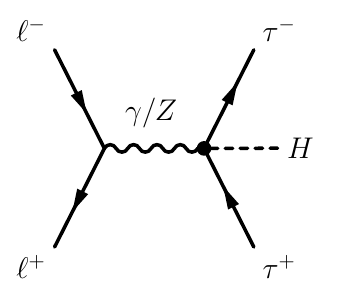}\\
\includegraphics[width=0.15\textwidth]{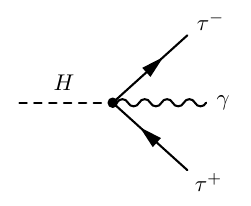}\qquad\qquad
\raisebox{0.1cm}{\includegraphics[width=0.17\textwidth]{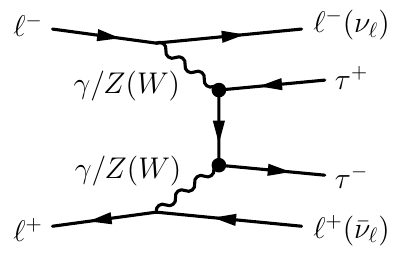}}
\caption{Processes sensitive to $\tau$ dipole moments at the FCC-ee and a muon collider (representative Feynman diagrams).} 
\label{fig:Feyn_Lett}
\end{figure}

Fig.~\ref{fig:Feyn_Lett} illustrates the relevant collider processes. The operators in Eq.~\eqref{eq:L_SMEFT} induce a $\tau^+\tau^-HV$ vertex ($V=\gamma,Z$), mediating both $H\to\tau^+\tau^-\gamma$ and $\ell^+\ell^-\to\tau^+\tau^-H$.

\section{III. Tau dipole moments at FCC-ee}

We now analyze the sensitivity to $\tau$ dipole moments at FCC-ee, 
see Fig.~\ref{fig:Feyn_Lett}, assuming the expected integrated luminosities 
reported in Table~\ref{tab:FCCee}. 
Our numerical analysis includes initial-state radiation (ISR), assumes an 80\% $\tau$-reconstruction efficiency, and imposes $5^\circ<\theta<175^\circ$ for detector acceptance \footnote{For further details on the analysis, see \cite{Buttazzo:2026ulg}.}. We decompose the $\ell^+\ell^-\to\tau^+\tau^-$ ($\ell=e,\mu$) cross section 
as $\sigma_{\ell\ell}=\sigma_{\ell\ell}^{\rm SM}+\sigma_{\ell\ell}^{\rm Lin}+\sigma_{\ell\ell}^{\rm Quad}$, where the last two terms denote SM--NP interference and the quadratic NP contribution, respectively.

Explicitly,
\begin{subequations}
\label{eq:sigma_ee_all}
\begin{align}
 &\!\sigma_{\ell\ell}^{\rm SM} \simeq \frac{1}{12\pi}\Biggl[
 \frac{e^4}{s} + \frac{2e^2g_Z^2g_V^2s\Delta}{D_Z}
 +\frac{g_Z^4s(g_V^2+g_A^2)^2}{D_Z}\Biggr],
 \label{eq:sigma_sm}\\
 &\!\sigma_{\ell\ell}^{\rm Lin} \simeq \frac{m_\tau v}{\sqrt{2}\pi s\Lambda^2}
 \operatorname{Re}\Biggl[e^3{\mathcal C}_{\tau\gamma}
 +\frac{eg_Z^2g_V^2{\mathcal C}_{\tau\gamma}\Delta s^2}{D_Z}
 \nonumber\\
 &\!\hspace{1.2cm}-\frac{g_Zg_V{\mathcal C}_{\tau Z}s^2
 \bigl[e^2\Delta+g_Z^2(g_V^2+g_A^2)\bigr]}{D_Z}\Biggr],
 \label{eq:sigma_lin}\\
 &\!\sigma_{\ell\ell}^{\rm Quad} \simeq \frac{v^2}{12\pi\Lambda^4}
 \Biggl[e^2|{\mathcal C}_{\tau\gamma}|^2
 +\frac{g_Z^2|{\mathcal C}_{\tau Z}|^2s^2(g_V^2+g_A^2)}{D_Z}
 \nonumber\\
 &\!\hspace{1.2cm}-\frac{2eg_Zg_V\operatorname{Re}
 ({\mathcal C}_{\tau\gamma}^*{\mathcal C}_{\tau Z})\Delta s^2}{D_Z}\Biggr].
 \label{eq:sigma_quad}
\end{align}
\end{subequations}
where $D_Z=\Delta^2s^2+M_Z^2\Gamma_Z^2$, $\Delta=1-M_Z^2/s$, $g_Z=g/c_W$, $g_V=-1/4+s_W^2$ and $g_A=-1/4$.
Eqs.~\eqref{eq:sigma_ee_all} neglect terms of order $m_\tau^2/s$, while retaining the chirality-suppressed interference proportional to $m_\tau v/s$ in Eq.~\eqref{eq:sigma_lin}, which dominates the sensitivity at the $Z$ pole but becomes less important at higher energies. The large Tera-Z luminosity probes sufficiently small dipole coefficients for the linear terms to dominate over the quadratic contributions.

\begin{table}[htb]
    \centering
    \scalebox{1.0}{
    \begin{tabular}{c|cccc}
    \hline
    Working point                     & $Z$ pole   ~~      & $WW$ ~~      & ~~$ZH$ & ~~$t\bar{t}$ \\ \hline
    $\sqrt{s}$ [GeV]                  & 87.9, 91.2, 94.3 & ~~ 157.5, 162.5 & ~~ 240 & ~~365        \\
    $\mathcal{L}$ [ab$^{-1}$] & 40, 125, 40      & 9.6, 9.6     & ~~ 10.8 &~~ 2.7        \\ \hline
    \end{tabular}%
    }
    \caption{Center of mass energies, $\sqrt{s}$, and integrated 
    luminosities, $\mathcal{L}$, at FCC-ee~\cite{FCC:2025lpp} 
    in the stages of a Tera-$Z$ ($Z$ pole), electroweak ($WW$), Higgs ($ZH$), and top ($t\bar t$) factories.}
    \label{tab:FCCee}
\end{table}

\begin{figure}[ht]
\includegraphics[width=0.37\textwidth]{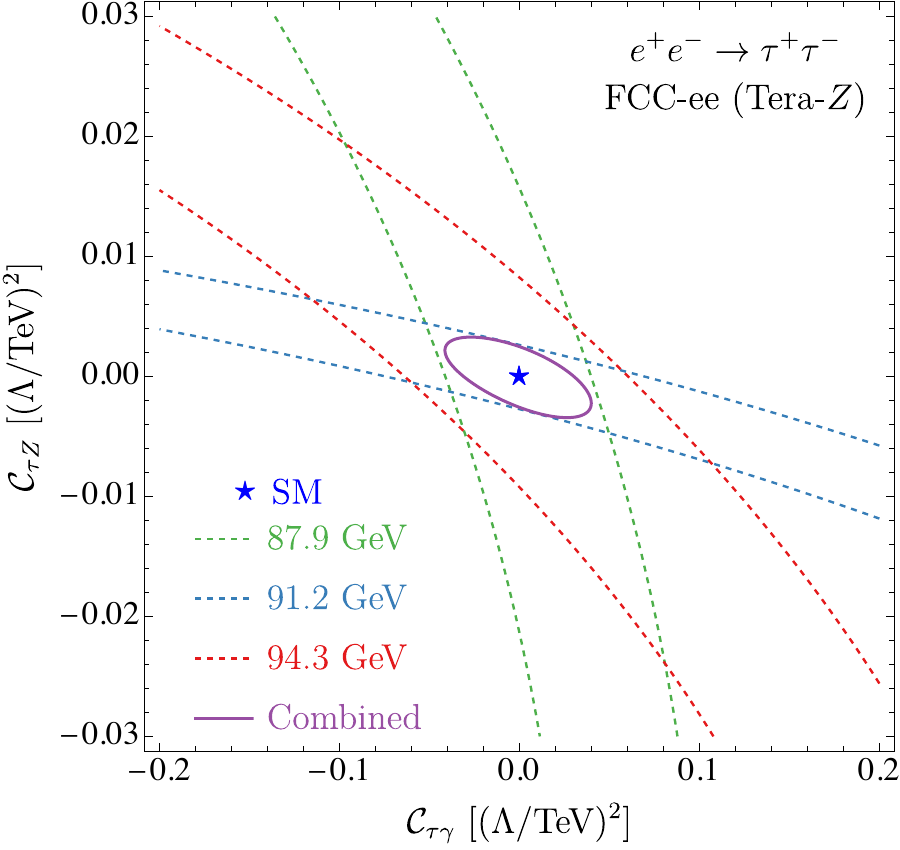}\\
\vspace{15pt}
\includegraphics[width=0.37\textwidth]{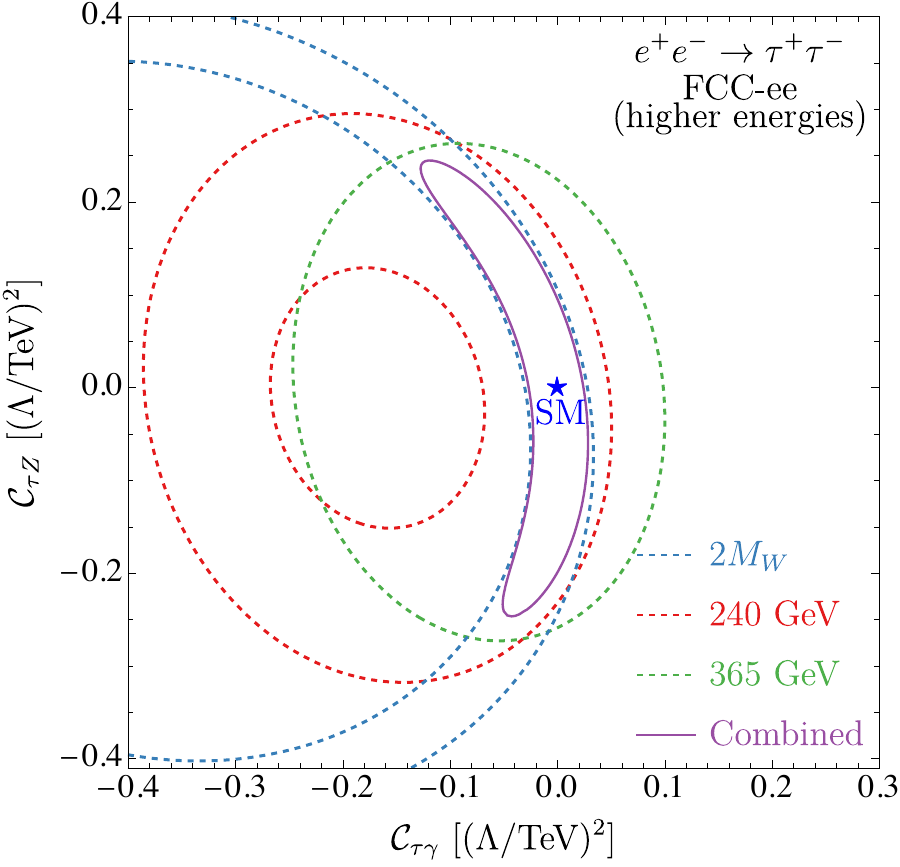}
\caption{95\% CL bounds in the $({\mathcal C}_{\tau \gamma}, {\mathcal C}_{\tau Z})$ plane at FCC-ee from $e^+e^- \!\to\! \tau^+\tau^-$ for the benchmark points of Table~\ref{tab:FCCee}. The bounds scale as $(\Lambda/{\rm TeV})^2$ and assume real ${\mathcal C}_{\tau \gamma}$ and ${\mathcal C}_{\tau Z}$.
} 
\label{fig:contours_FCCeeZ}
\end{figure}

We optimize the signal significance by imposing $m(\tau^+\tau^-)>0.8\sqrt{s}$, $\Delta R(\tau^+\tau^-)>0.4$, and $p_T(\tau^\pm)>0.1\sqrt{s}$. The resulting 95\% CL constraints in the $({\mathcal C}_{\tau\gamma},{\mathcal C}_{\tau Z})$ plane are shown in Fig.~\ref{fig:contours_FCCeeZ}.

At $\sqrt{s}\simeq M_Z$ (upper panel), linear NP contributions dominate the sensitivity, whereas quadratic effects become increasingly important at higher energies (lower panel). The factor $\Delta=1-M_Z^2/s$ in Eq.~\eqref{eq:sigma_lin} changes sign across the $Z$ pole, allowing the energy scan to probe complementary combinations of dipole coefficients and reduce degeneracies.
The resonant enhancement of $Z$ exchange makes the Tera-$Z$ stage particularly sensitive to the weak magnetic and electric dipole moments, $\Delta a_\tau^Z$ and $d_\tau^Z$.
Table~\ref{tab:constraints_eetata} reports the individual and combined 95\% CL bounds on $\Delta a_\tau^{(Z)}$ and $d_\tau^{(Z)}$. 

\begin{table}[]
    \centering
    \scalebox{1.0}{
    \begin{tabular}{ccccc}
    \hline
     $\sqrt{s}$   &   $|\Delta a_\tau|$   &  $|\Delta a_\tau^Z|$  &  $|d_\tau|$\,[$e \cdot {\rm cm}$] & $|d_\tau^Z|$\,[$e \cdot {\rm cm}$]\\
    \hline
    $M_Z$      &  $1.3\times 10^{-4}$ & $1.0\times 10^{-5}$ & $6.0\times 10^{-18}$ & $4.4\times 10^{-19}$\\
    $2M_W$     &  $1.2\times 10^{-4}$ & $9.8\times 10^{-4}$ & $3.3\times 10^{-18}$ & $3.6\times 10^{-18}$\\
    240 GeV    &  $2.8\times 10^{-4}$ & $9.3\times 10^{-4}$ & $3.1\times 10^{-18}$ & $4.4\times 10^{-18}$\\
    365 GeV    &  $1.0\times 10^{-3}$ & $1.0\times 10^{-3}$ & $3.5\times 10^{-18}$ & $5.5\times 10^{-18}$\\
    Combined   &  $8.2\times 10^{-5}$ & $1.0\times 10^{-5}$ & $2.5\times 10^{-18}$ & $4.4\times 10^{-19}$\\
    \hline
    \end{tabular}
    }
    \caption{95\% CL bounds on $\Delta a^{(Z)}_\tau$ and $d^{(Z)}_\tau$ at FCC-ee 
     from $e^+e^- \to \tau^+\tau^-$, for ${\mathcal C}_{\tau B}$ and ${\mathcal C}_{\tau W}$ purely real or imaginary.}
    \label{tab:constraints_eetata}
\end{table}



In the equivalent photon approximation, $\ell^+\ell^-\to\ell^+\ell^-\tau^+\tau^-$ is described by a convolution with $\hat\sigma_{\gamma\gamma}(\hat s)$:

\begin{equation}
\!\!\sigma_{\gamma\gamma}(s) \!=\! \!\int_{\xi_0}^1 \!{\rm d}\xi \int_{\xi}^1 \!\frac{{\rm d}x}{x} 
f_{\gamma/\ell}(x,Q^2) \, f_{\gamma/\ell}\!\Big(\frac{\xi}{x},Q^2\Big) 
\hat{\sigma}_{\gamma\gamma}(\hat{s})\,,
\label{eq:sigma_EPA}
\end{equation}
where $\xi=\hat{s}/s$, with $\sqrt{\hat{s}}$ the partonic center-of-mass energy, and $\xi_0=\hat{s}_{\rm min}/s$ fixed by the production threshold or kinematic cuts. We use the improved Weizs\"acker--Williams photon PDF $f_{\gamma/\ell}(x,Q^2)$~\cite{Frixione:1993yw}. 

Up to $1/\Lambda^4$, the partonic cross section reads $\hat\sigma_{\gamma\gamma}=\hat\sigma_{\gamma\gamma}^{\rm SM}+\hat\sigma_{\gamma\gamma}^{\rm Lin}+\hat\sigma_{\gamma\gamma}^{\rm Quad}$, with
\begin{align}
    &\hat{\sigma}_{\gamma\gamma}^{\rm SM} = \frac{e^4}{4 \pi \hat{s}} \Biggl[ (1+4\hat{\tau}-8\hat{\tau}^2) \log\left(\frac{1+\hat{\beta}}{1-\hat{\beta}}\right)- \hat{\beta}(1+4\hat{\tau}) \Biggr], 
    \nonumber\\
    &\hat{\sigma}_{\gamma\gamma}^{\rm Lin} = \frac{\sqrt{2} e^3 m_\tau v}{\pi \hat{s}} \frac{{\rm Re}\,{\mathcal C}_{\tau\gamma}}{\Lambda^2} \log\left(\frac{1+\hat{\beta}}{1-\hat{\beta}}\right), 
    \label{eq:gmgm2tata_xs_lin} 
    \\
    &\hat\sigma_{\gamma\gamma}^{\rm Quad}=\frac{e^2v^2}{\pi\Lambda^4}
    \Bigg[
    \left(
    ({\rm Re}\,\mathcal C_{\tau\gamma})^2-
    ({\rm Im}\,\mathcal C_{\tau\gamma})^2
    \right)
    \hat\tau
    \log\!\left(\frac{1+\hat\beta}{1-\hat\beta}\right)   \nonumber \\
    & \qquad \quad + 2|\mathcal C_{\tau\gamma}|^2\hat\beta  \Bigg], \nonumber
\end{align}
where $\hat\tau=m_\tau^2/\hat{s}$ and $\hat\beta=\sqrt{1-4m_\tau^2/\hat{s}}$. 
Unlike direct $\tau$ production, photon--photon collisions probe predominantly low-invariant-mass subprocesses, where SM--NP interference is comparatively enhanced. This linear contribution dominates the sensitivity, making this channel the most sensitive probe of $\Delta a_\tau$ at FCC-ee.

The DELPHI Collaboration studied $\tau$-pair production in photon-photon collisions at LEP-II using $\sim650\,\mathrm{pb}^{-1}$ at $\sqrt{s}=183$--$208\,\mathrm{GeV}$~\cite{DELPHI:2003nah}. We project yields using FCC-ee luminosities and DELPHI $\tau$-reconstruction efficiencies~\cite{DELPHI:2003nah}, as a benchmark for event rates and significance.

\begin{figure}[ht]
\includegraphics[width=0.37\textwidth]{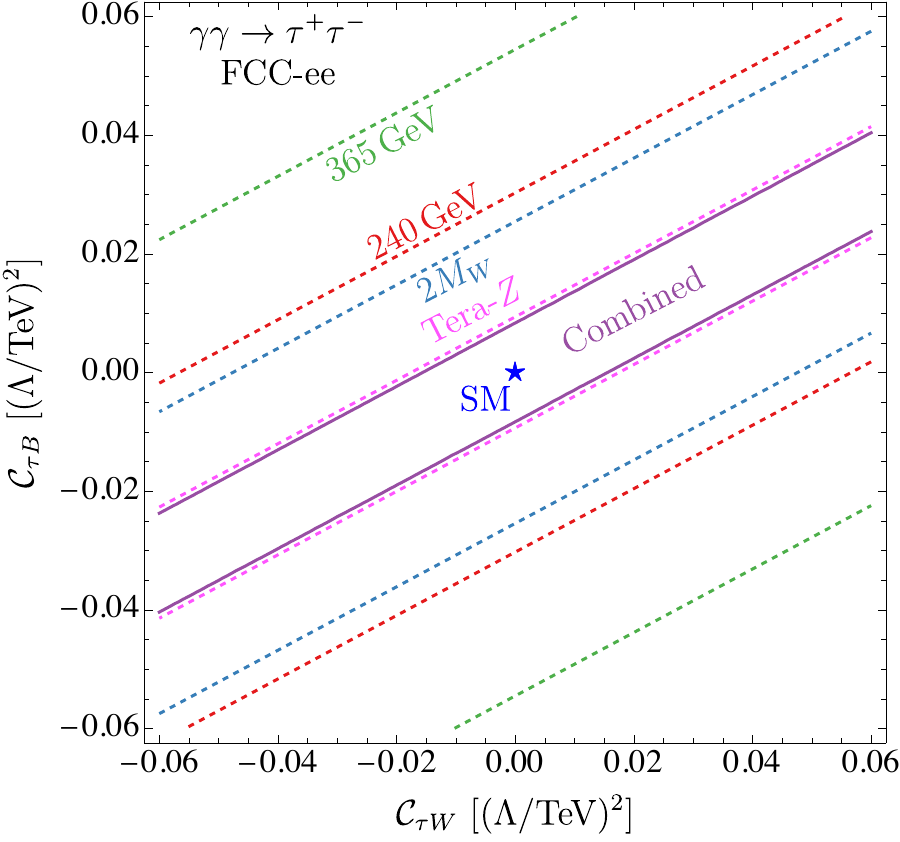}
\\[10pt]
\includegraphics[width=0.35\textwidth]{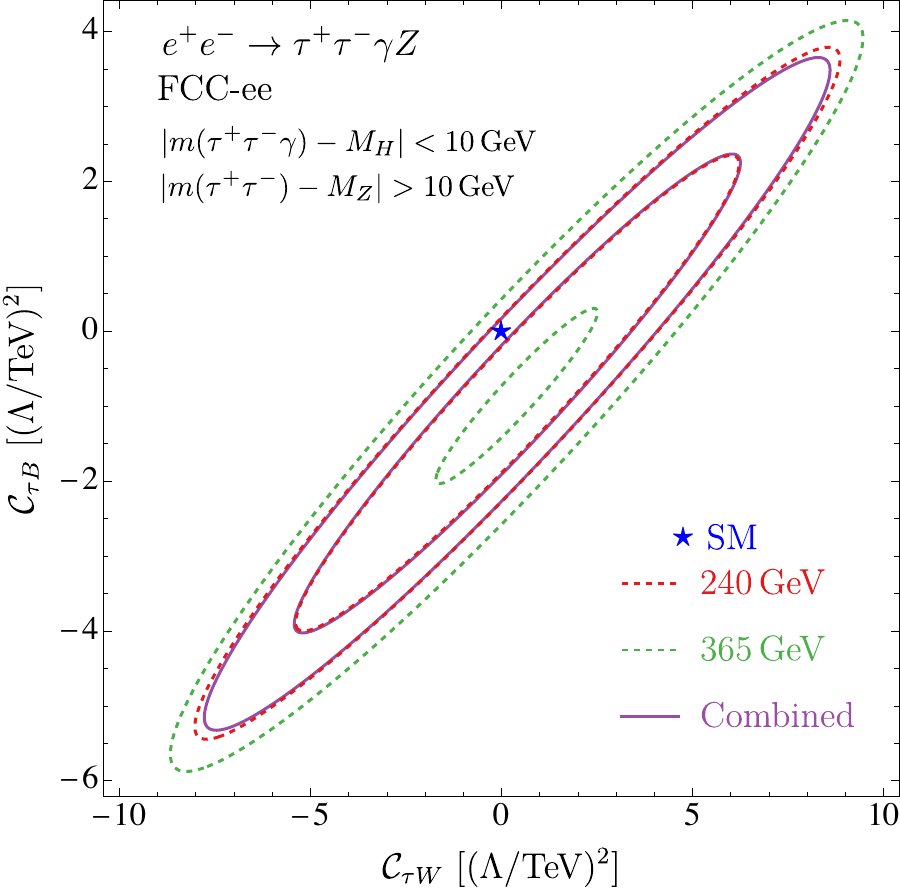}
\caption{95\% CL bounds in the $({\mathcal C}_{\tau W}, {\mathcal C}_{\tau B})$ plane at FCC-ee from $\gamma\gamma \to \tau^+\tau^-$ (top) and $H\to\tau^+\tau^-\gamma$ (bottom) for the benchmark points of Table~\ref{tab:FCCee}. The bounds scale as $(\Lambda/{\rm TeV})^2$ and assume real ${\mathcal C}_{\tau W}$ and ${\mathcal C}_{\tau B}$.}
\label{fig:contours_ZF_real}
\end{figure}

The top panel of Fig.~\ref{fig:contours_ZF_real} shows the 95\% CL bounds from $\gamma\gamma\to\tau^+\tau^-$ at FCC-ee. 
The Tera-$Z$ program provides the strongest constraints on ${\mathcal C}_{\tau\gamma}$, benefiting from its unprecedented integrated luminosity.
Table~\ref{tab:constraints_gmgm} gives $|\Delta a_\tau|\lesssim3.0\times10^{-5}$, almost three times stronger than the individual electromagnetic bound from direct $\tau^+\tau^-$ pair production. 
However, in UV scenarios with correlated electromagnetic and weak dipoles of comparable size, the more stringent bound on $\Delta a_\tau^Z$ from $e^+e^-\to\tau^+\tau^-$ can yield an even stronger indirect constraint on $\Delta a_\tau$.

\begin{table}[ht]
    \centering
    \scalebox{0.82}{
    \begin{tabular}{cccccc}
    \hline
     $\sqrt{s}$   &   $M_Z$  &   $2M_W$   &   240 GeV   &   365 GeV  & Combined  \\
    \hline
    $|\Delta a_\tau|$    & $3.4\times 10^{-5}$         &  $9.2\times 10^{-5}$ & $1.1\times 10^{-4}$ & $2.0\times 10^{-4}$ & $3.0\times 10^{-5}$ \\
    $|d_\tau|$\,[$e \cdot {\rm cm}$]    &  $1.9\times 10^{-17}$  &  $2.9\times 10^{-17}$ & $3.0\times 10^{-17}$ & $3.8\times 10^{-17}$ & $1.7\times 10^{-17}$ \\
    \hline
    \end{tabular}
    }
    \caption{95\% CL bounds on $\Delta a_\tau$ and $d_\tau$ at FCC-ee from 
$\gamma\gamma \to \tau^+\tau^-$, for ${\mathcal C}_{\tau B}$ and ${\mathcal C}_{\tau W}$ purely real or imaginary. 
} 
    \label{tab:constraints_gmgm}
\end{table}


We also consider $H\to\tau^+\tau^-\gamma$ at FCC-ee through $e^+e^-\to HZ\to\tau^+\tau^-\gamma Z$. The dipole operators in Eq.~\eqref{eq:L_SMEFT} contribute to its decay width as~\cite{Buttazzo:2020ibd}

\begin{equation}
\!\!\Gamma(H \to \tau^+\tau^-\gamma)_{\rm NP} =
\frac{e y_\tau M_H^3}{64\pi^3}
\frac{{\rm Re}\, {\mathcal C}_{\tau \gamma}}{\Lambda^2}
+ \frac{M_H^5}{768\pi^3}
\frac{|{\mathcal C}_{\tau \gamma}|^2}{\Lambda^4}.
\label{eq:radiative_width}
\end{equation}
To isolate the signal, we impose a Higgs-mass window $|m(\tau^+\tau^-\gamma)-M_H|<10\,\mathrm{GeV}$, a $Z$ veto $|m(\tau^+\tau^-)-M_Z|>10\,\mathrm{GeV}$ to suppress the dominant on-shell $Z$ background, and an additional cut $p_T>M_H/10$ to optimize the signal sensitivity.
The resulting 95\% CL contours in the $({\mathcal C}_{\tau W},{\mathcal C}_{\tau B})$ plane are shown in the bottom panel of Fig.~\ref{fig:contours_ZF_real}, with the corresponding bounds on $\Delta a_\tau^{(Z)}$ and $d_\tau^{(Z)}$ in Table~\ref{tab:constraints_eettaz}. 
Though weaker than pair-production bounds (Tables~\ref{tab:constraints_eetata} and~\ref{tab:constraints_gmgm}), this decay could improve current LHC limits.

\begin{table}[]
    \centering
    \scalebox{1.0}{
    \begin{tabular}{ccccccc}
    \hline
     $\sqrt{s}$   &   $|\Delta a_\tau|$  &   $|\Delta a_\tau^Z|$   &  $|d_\tau|$\,[$e \cdot {\rm cm}$] & $|d_\tau^Z|$\,[$e \cdot {\rm cm}$]\\
    \hline
    240 GeV    &  $7.5\times 10^{-4}$ & $2.0\times 10^{-2}$ & $1.3\times 10^{-17}$ & $1.1\times 10^{-16}$\\
    365 GeV    &  $2.6\times 10^{-3}$ & $2.9\times 10^{-2}$ & $2.1\times 10^{-17}$ & $1.6\times 10^{-16}$\\
    Combined   & $7.0\times 10^{-4}$ & $1.9\times 10^{-2}$  & $1.2\times 10^{-17}$ & $1.1\times 10^{-16}$\\
    \hline
    \end{tabular}
    }
    \caption{95\% CL bounds on $\Delta a^{(Z)}_\tau$ and $d^{(Z)}_\tau$ at FCC-ee from $H \to \tau^+\tau^-\gamma$, for ${\mathcal C}_{\tau B}$ and ${\mathcal C}_{\tau W}$ purely real or imaginary.}
    \label{tab:constraints_eettaz}
\end{table}


\section{IV. Tau dipole moments at a \boldmath{$\mu$}C}
\begin{figure}[b]
\includegraphics[width=0.45\textwidth]{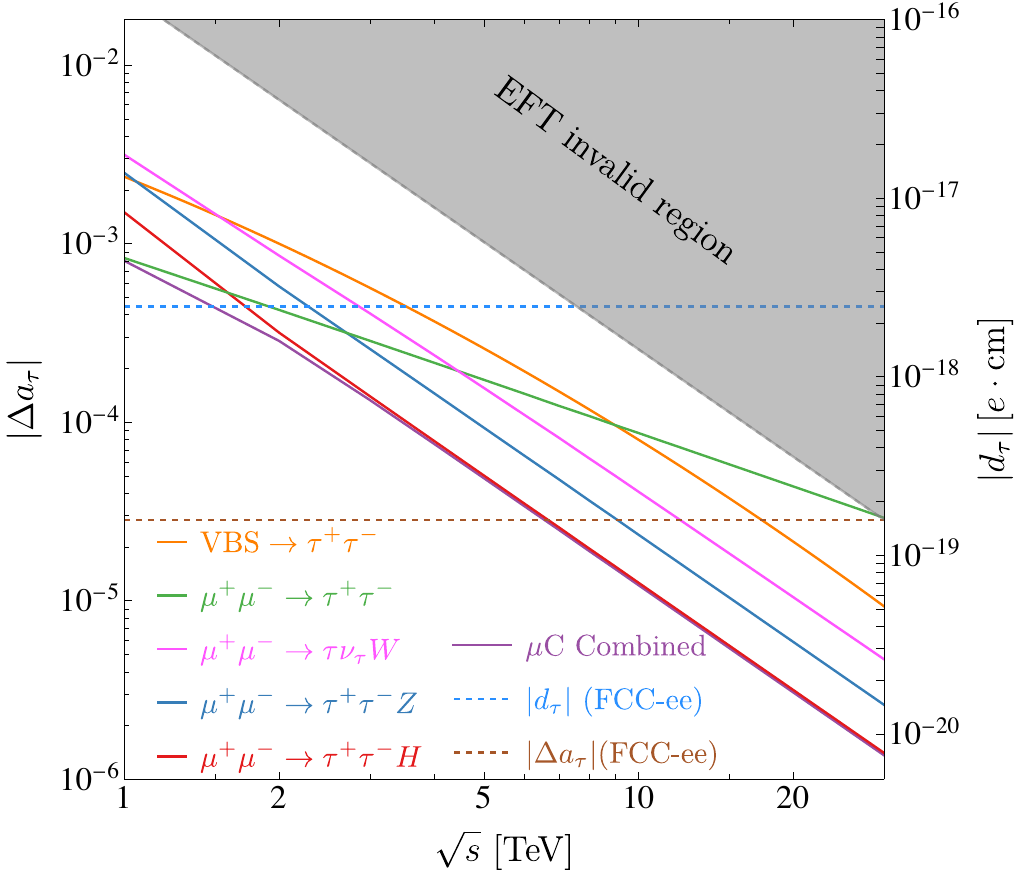}
\caption{95\% CL bounds on $\Delta a_\tau$ and $d_\tau$ at multi-TeV muon collider as a function of $\sqrt{s}$, assuming real ${\mathcal C}_{\tau B}$ and ${\mathcal C}_{\tau W}$.
The grey region denotes perturbative unitarity violation in Eq.~\eqref{eq:perturbative}.}
\label{fig:muC_constraint}
\end{figure}

For a $\mu$C we assume $\mathcal L=10\,\mathrm{ab}^{-1}(\sqrt{s}/10\,\mathrm{TeV})^2$ at $\sqrt{s}=3,6,10,14\,\mathrm{TeV}$. We adopt the FCC-ee simulation setup, adding $10^\circ<\theta<170^\circ$ to mitigate beam-induced background. 
The numerical projections are obtained at fixed dipole Wilson coefficients, neglecting their renormalization-group evolution. The resulting running effects are expected to be subleading within the accuracy of the present sensitivity estimates~\cite{Buttazzo:2026jfu}. 
We first consider $\mu^+\mu^-\to\tau^+\tau^-$, whose sensitivity to dipole interactions is dominated by quadratic NP contributions at multi-TeV energies. Figure~\ref{fig:muC_constraint} shows the resulting 95\% CL bounds on $\Delta a_\tau$ and $d_\tau$.
With $\mathcal L\propto s$, the bounds improve as $1/\sqrt{s}$ because the SM background falls as $1/s$, 
eventually approaching the perturbative-unitarity limit. The bounds on $\Delta a_\tau$ remain weaker than the FCC-ee limit, 
even at $\sqrt{s}\sim20\,\mathrm{TeV}$. By contrast, quadratic $d_\tau$ contributions benefit strongly from the higher $\mu$C energies.

In the high-energy limit $s\gg M_{H,Z}^2$, the process $\mu^+\mu^-\to\tau^+\tau^-H$ has the differential cross section

\begin{align}
&\!\!\frac{d\sigma_{\tau\tau H}}{dx_1 dx_2}
\simeq \frac{1}{192\pi^3}\frac{s}{\Lambda^4}
(1-x_1-x_2+2x_1x_2) \cdot \nonumber
\\
&\!\!\!\left[ |{\mathcal C}_{\tau \gamma}|^2 e^2 \!\!-\! 2 e g_Z g_V {\rm Re}({\mathcal C}^*_{\tau Z} {\mathcal C}_{\tau \gamma}) \!+\! |{\mathcal C}_{\tau Z}|^2 g_Z^2 (g_A^2 \!+\! g_V^2)
\right],
\label{eq:mm2tth}
\end{align}
where $x_i=2Q\!\cdot\!k_i/Q^2$, with $k_i$ denoting the final-state $\tau$ four-momenta and $Q$ the total incoming four-momentum. Besides the pair-production cuts, we require $m(\tau^+\tau^-)>\sqrt{s}/10$ to suppress the SM background from $\mu^+\mu^-\to ZH$, with $Z\to\tau^+\tau^-$. We tag $H\to b\bar b$ with 80\% efficiency and assume a 15\% misidentification rate for $Z\to b\bar b$ as $H\to b\bar b$.

At high energies, $\mu^+\mu^-\to\tau^+\tau^-Z$ and $\mu^+\mu^-\to\tau^\pm\nu_\tau W^\mp$ exhibit analogous NP effects through longitudinal bosons, consistent with the Goldstone equivalence theorem~\cite{Cornwall:1974km,Lee:1977eg,Chanowitz:1985hj}. The $s$ growth in Eq.~\eqref{eq:mm2tth} and $\mathcal L\propto s$ yield dipole bounds scaling as $1/s$ (Fig.~\ref{fig:muC_constraint}). Within these scaling assumptions, associated production retains a more favorable energy dependence than pair production.
The resulting bounds on $\Delta a_\tau$ and $d_\tau$ are comparable to those from $\mu^+\mu^-\to\tau^+\tau^-$ at $\sqrt{s}\sim2\,\mathrm{TeV}$ and about an order of magnitude stronger at $\sqrt{s}\sim10\,\mathrm{TeV}$.

Unlike at FCC-ee, probing $\tau$ dipole moments through VBS $\tau^+\tau^-$ production in the linear NP regime is challenging at a $\mu$C due to large QCD dijet backgrounds~\cite{Han:2021kes}. To isolate VBS from muon annihilation and maximize NP sensitivity, we impose $p_T(\tau^\pm)>0.1x\sqrt{s}$ and $x\sqrt{s}<m(\tau^+\tau^-)<0.8\sqrt{s}$, optimizing $x$ at each $\sqrt{s}$ (typically $x=0.2$--$0.3$). 
At high energies, the NP signal is dominated by $\mu^+\mu^-\to\tau^+\tau^-Z$ ($Z\to\nu\bar\nu$), yielding a projected sensitivity to $\Delta a_\tau$ that improves as $1/s$.

$W^+W^-$ fusion produces $\mathcal O(10^7)$ Higgs bosons at a $\sqrt{s}\sim10\,\mathrm{TeV}$ $\mu$C. We simulate the full processes $\mu^+\mu^-\to\tau^+\tau^-\gamma+\mu^+\mu^-/\nu\bar\nu$ to capture $H\to\tau^+\tau^-\gamma$. We apply the FCC-ee Higgs-mass and $Z$-veto cuts, together with $p_T > 50$ GeV for all particles to suppress non-resonant backgrounds. 
At $\sqrt{s}=14\,\mathrm{TeV}$, the projected sensitivity reaches $|\Delta a_\tau|\approx6\times10^{-4}$, weaker than that of the other channels considered but comparable to the FCC-ee reach from radiative Higgs decays.

\begin{figure}[t]
\includegraphics[width=0.49\textwidth]{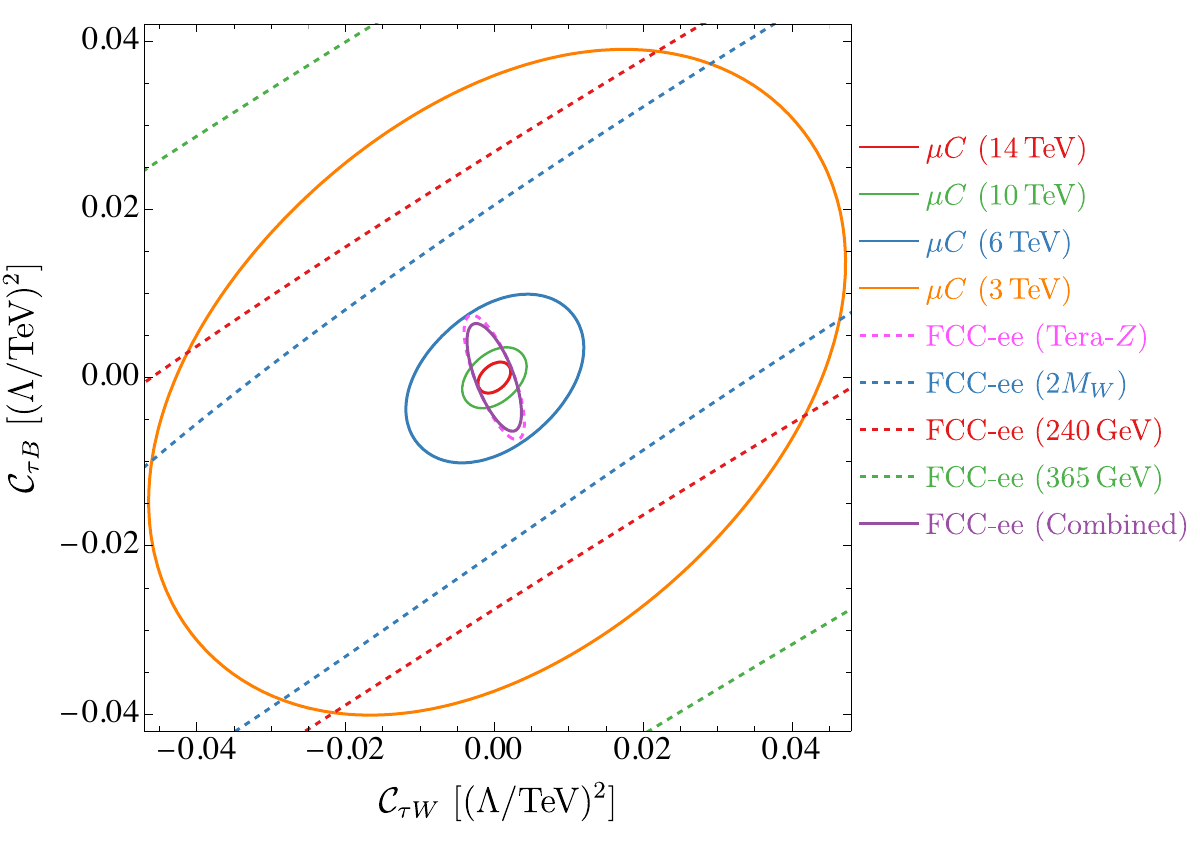}
\caption{95\% CL bounds in the $({\mathcal C}_{\tau W},\,{\mathcal C}_{\tau B})$ plane at FCC-ee and multi-TeV muon collider. The bounds scale as $(\Lambda/{\rm TeV})^2$ and assume real ${\mathcal C}_{\tau W}$ and ${\mathcal C}_{\tau B}$.
} \label{fig:contour_all}
\end{figure}

Fig.~\ref{fig:contour_all} compares the 95\% CL constraints from FCC-ee and the $\mu$C in the $({\mathcal C}_{\tau W},{\mathcal C}_{\tau B})$ plane. While the Tera-$Z$ program provides stringent bounds on the weak dipole moments, $\Delta a_\tau^Z$ and $d_\tau^Z$, the $\mu$C offers greater sensitivity to both electromagnetic and weak dipoles above $\sqrt{s}\sim10\,\mathrm{TeV}$. At 14 TeV, the projected bound on $d_\tau$ improves by over two orders of magnitude relative to FCC-ee.

\section{V. CONCLUSIONS}
At FCC-ee, the Tera-$Z$ luminosity makes SM--NP interference dominant in direct $e^+e^-\to\tau^+\tau^-$ production, unlike in the quadratic-dominated LEP-II regime. The energy scan across the $Z$ pole exploits the sign-changing part of this interference to resolve complementary combinations of real dipole coefficients. Photon collisions instead probe low-invariant-mass subprocesses, in which linear NP effects are comparatively enhanced; combined with the Tera-$Z$ luminosity, they give the strongest individual electromagnetic bound, $|\Delta a_\tau|\lesssim3.0\times10^{-5}$ (Table~\ref{tab:constraints_gmgm}). Direct production reaches $|\Delta a_\tau^Z|\lesssim1.0\times10^{-5}$ and can be equally or more incisive in NP scenarios with correlated electromagnetic and weak dipoles. It also gives the strongest projected bounds $|d_\tau|\lesssim2.5\times10^{-18}\,e\,\mathrm{cm}$ and $|d_\tau^Z|\lesssim4.4\times10^{-19}\,e\,\mathrm{cm}$ (Table~\ref{tab:constraints_eetata}).

At a multi-TeV muon collider, the energy growth of associated Higgs production $\mu^+\mu^-\to\tau^+\tau^-H$ provides a distinct mechanism, with sensitivity improving as $1/s$ under the assumed luminosity scaling. At $\sqrt{s}\sim14\,\mathrm{TeV}$, the projected bounds on $\Delta a_\tau$ and $d_\tau$ improve by roughly one and two orders of magnitude relative to FCC-ee (Fig.~\ref{fig:muC_constraint}). 
These complementary probes extend the $\tau$-dipole reach by several orders of magnitude over current bounds. Assessing systematic uncertainties is essential to establish achievable sensitivities.

\medskip
\vskip 0.1cm
\textbf{Acknowledgments:}
We thank B. Batell, I. Brivio, C. Degrande, G. Durieux, T. Han, M. Hoferichter, F. Jaffredo, J. Liu, O. Mattelaer, M. Passera, and X. Wang for useful discussions. 
This research is partially supported by the IISN convention 4.4517.08, ``Theory of fundamental interactions.'' Y. Ma acknowledges the support as a Postdoctoral Fellow of the Fond de la Recherche Scientifique de Belgique (F.R.S.-FNRS), Belgium.
Computational resources have been provided by the supercomputing facilities of the Universit\'e catholique de Louvain (CISM/UCL) and the Consortium des \'Equipements de Calcul Intensif en 
F\'ed\'eration Wallonie Bruxelles (C\'ECI) funded by the Fond de la Recherche Scientifique de Belgique (F.R.S.-FNRS) under convention 2.5020.11 and by the Walloon Region.
G. Levati gratefully acknowledges financial support by the Swiss National Science Foundation (Project No.\
TMCG-2\_213690). Z.Q. Wang acknowledges support from the China Scholarship Council (CSC)- UCLouvain Co‑funding PhD Fellowship (Grant No.\ 202206040017).
P. Paradisi received funding by the European Union’s Horizon 2020 research and innovation program under the Marie Sklodowska-Curie grant agreements n.~860881~-~HIDDeN, n.~101086085~-~ASYMMETRY, by the Italian MUR Departments of Excellence grant 2023-2027 ``Quantum Frontiers'' and by the European Union - Next Generation EU and by the Italian Ministry of University and Research (MUR) via the PRIN 2022 project n. 2022K4B58X - AxionOrigins. D. Buttazzo is partially funded by the European Union - Next Generation EU and MUR through the grant PRIN 202289JEW4. We also acknowledge support from the COMETA COST Action CA22130.\\


\bibliographystyle{utphys}
\bibliography{ref.bib}

\end{document}